\documentclass[useAMS,usenatbib]{mn2e}
\usepackage{graphicx}
\usepackage{amssymb}
\usepackage{amsmath}
\usepackage{verbatim}
\title[Lightcurves of Stars \& Exoplanets]{Lightcurves of Stars \& Exoplanets:\\ Estimating Inclination, Obliquity, and Albedo}
\author[N.B.~Cowan, P.A.~Fuentes and H.M.~Haggard]{Nicolas~B.~Cowan$^{1,2}$, Pablo~A.~Fuentes$^{3}$ and Hal~M.~Haggard$^4$\\
$^1$Center for Interdisciplinary Exploration and Research in Astrophysics (CIERA),\\ Northwestern University, 2131 Tech Dr.\, Evanston, IL 60208, USA (n-cowan@northwestern.edu)\\ 
$^2$Department of Physics and Astronomy, Northwestern University, 2145 Sheridan Rd., F165, Evanston, IL 60208, USA\\
$^3$Department of Astronomy, University of Chile, Camino El Observatorio \# 1515, Las Condes, Santiago, Chile\\
$^4$Centre de Physique Th\'eorique de Luminy, Campus de Luminy, Case 907 13288 Marseille cedex 9, France}

\begin{onecolumn}
\begin{document}

\maketitle

\begin{abstract}
Distant stars and planets will remain spatially unresolved for the foreseeable future. It is nonetheless possible to infer aspects of their brightness markings and viewing geometries by analyzing disk-integrated rotational and orbital brightness variations.  We compute the harmonic lightcurves, $F_l^m(t)$, resulting from spherical harmonic maps of intensity or albedo, $Y_l^m(\theta,\phi)$, where $l$ and $m$ are the total and longitudinal order.  It has long been known that many non-zero maps have no lightcurve signature, e.g., odd $l>1$ belong to the nullspace of harmonic thermal lightcurves.  We show that the remaining harmonic lightcurves exhibit a predictable inclination-dependence. Notably, odd $m>1$ are present in an inclined lightcurve, but not seen by an equatorial observer.  We therefore suggest that the Fourier spectrum of a thermal lightcurve may be sufficient to determine the orbital inclination of non-transiting short-period planets, the rotational inclination of stars and brown dwarfs, and the obliquity of directly imaged planets. In the best-case scenario of a nearly edge-on geometry, measuring the $m=3$ mode of a star's rotational lightcurve to within a factor of two provides an inclination estimate good to $\pm 6^{\circ}$, assuming stars have randomly distributed spots.  Alternatively, if stars have brightness maps perfectly symmetric about the equator, their lightcurves will have no $m=3$ power, regardless of orientation.  In general, inclination estimates will remain qualitative until detailed hydrodynamic simulations and/or occultation maps can be used as a calibrator. We further derive harmonic reflected lightcurves for tidally-locked planets; these are higher-order versions of the well-known Lambert phase curve. We show that a non-uniform planet may have an apparent albedo 25\% lower than its intrinsic albedo, even if it exhibits precisely Lambertian phase variations.  Lastly, we provide low-order analytic expressions for harmonic lightcurves that can be used for fitting observed photometry; as a general rule, edge-on solutions cannot simply be scaled by $\sin i$ to mimic inclined lightcurves.     
\end{abstract}

\section{Introduction}
\subsection{Motivation}
Extrasolar planets are sufficiently small and distant that they will remain spatially unresolved for the foreseeable future.  It is nonetheless possible to infer spatial inhomogeneities on these bodies through a) occultations, as when a planet passes behind its host star \citep[][]{Majeau_2012, deWit_2012}, or b) orbital and rotational motion \citep[e.g.,][]{Knutson_2007, Cowan_2009}.  Exoplanets are only the most recent astronomical objects amenable to such methods, after stars, minor planets, and accretion disks.  Therefore, while we will often refer in this paper to ``planets,'' it should be understood that the same formalism can be applied to any spherical body.    

In the current study we consider photometric variability due to rotational and orbital motion, i.e. method b).  We seek analytic expressions for the time-variations in disk-integrated brightness measured by a distant observer, as a function of the intrinsic spatial inhomogeneities of the planet and the system geometry.  In particular, we consider changes in disk-integrated thermal flux due to spatial inhomogeneities in thermal emission, and variations in disk-integrated reflectance due to spatial inhomogeneities in albedo.  

In addition to betraying brightness markings on stars and planets, rotational and orbital phase variations have the potential to constrain viewing geometry. Possible applications include: the thermal phase variations of non-transiting hot Jupiters might hint at their orbital inclination, breaking the $M\sin{i}$ degeneracy and allowing for improved mass estimates; the rotational phase variations of a transiting planet's host star may be sufficient to infer its rotational inclination (a.k.a. stellar obliquity), a useful discriminator between planet migration scenarios \citep{Winn_2005}; the rotational photometric variations of a directly-imaged planet might encode information about its rotational inclination which ---when combined with the astrometrically inferred orbital inclination--- provides an estimate of planetary obliquity, telling us about planet formation \citep[][]{Tremaine_1991}.

\subsection{Forward vs.\ Inverse Problem}
Inferring the properties of a star or planet based on its disk-integrated brightness is an \emph{inverse problem}, as opposed to the \emph{forward problem} of predicting the photometry of an object based on its properties. We approximate the forward problem as linear in the planet map, $M(\theta, \phi)$:
\begin{equation} \label{convolution}
F(t) = \oint K(\theta, \phi, t) M(\theta, \phi) d\Omega, 
\end{equation}    
where $F(t)$ is the observed flux, $K(\theta, \phi, t)$ is the kernel, $\theta$ and $\phi$ are planetary co-latitude and longitude, respectively, and $d\Omega = \sin\theta d\theta d\phi$.  As we will see below, the kernel is non-negative and unimodal.  It is therefore tempting to think of \eqref{convolution} as a convolution, and the inverse problem as a deconvolution. For thermal lightcurves, $K$ has a fixed shape and this description is formally correct; in other cases it is merely a useful analogy.

The inverse problem, solving for $M$ given $K$ and $F$, is a Fredholm integral equation of the first kind and is non-trivial \citep[][]{Aster_2013}. Inverse problems are typically under-constrained, and ``exo-cartography'' is no exception.  First of all, there are non-zero maps that produce flat lightcurves, a so-called nullspace.\footnote{The term ``kernel'' is often used interchangeably with ``nullspace'' in mathematical physics, but we eschew that terminology here because ``kernel'' already has a central role in convolutions.} Secondly, even non-zero harmonic lightcurves are sometimes proportional to each other. This is not surprising, since linear transformations need not preserve angles: planetary maps that are orthogonal are not necessarily transformed to lightcurves that are orthogonal. The bottom line is that attempts to map the brightness markings of distant objects suffer from formal degeneracies, even in the limit of noiseless observations.

If the orientation of the planet or star is not known \emph{a priori}, then the problem can be expressed as
\begin{equation}
F(t) = \oint K(\mathbb{G}, \theta, \phi, t) M(\theta, \phi) d\Omega, 
\end{equation}  
where $\mathbb{G}$ represents the unknown geometry (e.g., inclination or obliquity). The object is then to solve for $\mathbb{G}$ and $M(\theta, \phi)$ knowing $F$ and \emph{the form} of $K$. It has been demonstrated in numerical experiments, for example, that one can simultaneously constrain a planet's two-dimensional albedo map, obliquity and obliquity phase \citep[][]{Kawahara_2010, Kawahara_2011, Fujii_2012}.   

\subsection{Harmonic Lightcurves}\label{harmonic_lightcurves} 
In order to develop an analytic solution to \eqref{convolution}, it is necessary to express the planetary map analytically.  In general this is done by decomposing $M$ using an orthonormal basis.  The obvious basis maps for a spherical planet are spherical harmonics.   Any continuous, static albedo map, $M(\theta,\phi)$, may be decomposed as\\
\begin{minipage}[t]{0.45\linewidth}
\begin{equation}\label{map_construction}
M(\theta, \phi) = \sum_{l=0}^{\infty} \sum_{m=-l}^{l} C_{l}^m Y_{l}^{m}(\theta, \phi),
\end{equation}  
\end{minipage}
\begin{minipage}[t]{0.45\linewidth}
\begin{equation}\label{coefficients}
C_{l}^{m} = \frac{1}{4\pi}\oint M(\theta, \phi) Y_{l}^{m}(\theta, \phi) d\Omega.
\end{equation}
\end{minipage}

The real spherical harmonics are given by
\begin{equation}
Y_{l}^{m}(\theta, \phi) = \left\{ \begin{array}{ll}
 N_{l}^{m} P_{lm}(\cos\theta) \cos(m\phi) & \textrm{if $m\ge0$}\\
N_{l}^{|m|} P_{l|m|}(\cos\theta) \sin(|m|\phi) & \textrm{if $m<0$,} \end{array} \right.
\end{equation} 

where $P_{lm}$ is the associated Legendre polynomial without the Condon-Shortley phase, $(-1)^m$.

We adopt the geodesy normalization (unit power) for real spherical harmonics,\\
\begin{minipage}[t]{0.45\linewidth}
\begin{equation}
N_{l}^{m} = \left\{ \begin{array}{ll}
1 & \textrm{if $l=0$}\\
\sqrt{\frac{2(2l+1)(l-m)!}{(l+m)!}} & \textrm{if $l>0$,} \end{array} \right.
\end{equation}
\end{minipage}
\begin{minipage}[t]{0.45\linewidth}
\begin{equation}
\frac{1}{4\pi}\oint Y_l^m(\theta, \phi) Y_\lambda^\mu(\theta, \phi) d\Omega = \delta_{l\lambda}\delta_{m\mu}.
\end{equation}
\end{minipage}

The lightcurve signature of a spherical harmonic, or \emph{harmonic lightcurve}, is 
\begin{equation}\label{harmonic_lightcurve}
F_{l}^{m}(t) = \oint K(\theta, \phi, t) Y_{l}^{m}(\theta, \phi) d\Omega.
\end{equation} 
It is perfectly equivalent to think of this as decomposing the kernel into spherical harmonics. Aside from the current application of photometric variability which dates to \cite{Russell_1906}, this sort of formalism has broad applications throughout astrophysics \citep[e.g., constraining $\vec{B}$-field morphology of Ap stars via harmonic analysis of time-variable spectra;][]{Deutsch_1958, Deutsch_1970}.

In this paper we present harmonic lightcurves for a few cases of immediate interest. We tackle thermal lightcurves in \S\ref{thermal_lightcurves} and address the more complex case of reflected lightcurves in \S\ref{reflected_lightcurves}.  In both of those sections we begin by describing our model assumptions, then present solutions to special cases before moving on to the general solution. Whenever possible, we solve the integrals analytically by hand and/or with \emph{Mathematica}.  When symbolic solutions are too messy to have intuitive value, we use \emph{IDL} to compute and plot numerical integrals. We discuss possible applications and implications of this work in \S\ref{discussion}.

\section{Thermal Lightcurves}\label{thermal_lightcurves}
\subsection{Model Formalism}
We assume a spherical planet, static brightness map, diffuse thermal emission, and neglect limb-darkening. The requirement of a static map depends on context.  For mapping star spots or patchy clouds on a brown dwarf, the rotation period is the relevant timescale. When mapping the diurnal heating pattern of a planet, on the other hand, one requires stability on the orbital period \citep[for more about the various sources of planetary thermal variability see][]{Cowan_2012c}.   

The flux, $F$, in this case is the disk-integrated thermal flux from the planet. The kernel is proportional to the visibility of a given region of the planet at time $t$: $K(\theta, \phi, t) = \frac{1}{\pi}V(\theta, \phi, t)$,  where the visibility, $V$, is unity at the sub-observer location, drops as the cosine of the angle from the sub-observer location, $\gamma_o$, and is zero on the far side of the planet:
\begin{equation}\label{V}
V(\theta, \phi, t) = \max[\cos\gamma_o, 0] = \max[\sin\theta\sin\theta_{o}\cos(\phi-\phi_{o})+\cos\theta\cos\theta_{o}, 0],
\end{equation}
where $\theta_{o}$ and $\phi_o$ are the sub-observer co-latitude and longitude, respectively. The piece-wise defined kernel leads to much of the difficulty in solving the forward problem analytically.

The entire time-dependence of the forward problem comes in through the sub-observer position. In the absence of precession, the sub-observer co-latitude is constant, $\theta_{o}(t) = \theta_{o}$. The sub-observer longitude decreases linearly with time (we define longitude increasing to the East, with the planet rotating from West to East): $\phi_{o}(t) = \phi_{o}(0) - \omega_{\rm rot} t$, where $\omega_{\rm rot}$ is the rotational angular frequency \emph{in an inertial frame} (e.g., $\omega_{\rm rot} = 2\pi/23.93$~hr$^{-1}$ for Earth).

The thermal harmonic lightcurves are given by
\begin{equation}\label{thermal_harmonic_lightcurve}
F_{l}^m(t) = \frac{1}{\pi}\oint V(\theta, \phi, t) Y_{l}^{m}(\theta, \phi) d\Omega.
\end{equation}

Integrating the piece-wise defined kernel over the entire sphere is equivalent to integrating the non-zero part of the kernel, $K_{\rm nz}(\theta , \phi , t) = \frac{1}{\pi}(\sin \theta \sin \theta_{o} \cos ( \phi - \phi_{o}) + \cos \theta \cos \theta_{o}$), over the visible hemisphere.  The limits of integration are then defined by the limb, the locus of points with $\gamma_o = \pi/2$. From \eqref{V}, the limb satisfies:
\begin{equation} \label{limb}
\tan \theta_{\rm limb}  = \frac {-1}{\tan \theta_{o}\cos (\phi - \phi_{o}) },
\end{equation} 
as shown in Figure~\ref{vis_map}.

For a planet viewed equator-on ($\theta_{o} = \pi/2$) the kernel simplifies to: $K_{\rm nz}(\theta , \phi , t) = \frac{1}{\pi}\sin \theta \cos ( \phi - \phi_{o})$.

\subsection{Equator-On Thermal Lightcurve}
\begin{figure}
\includegraphics[width=85mm]{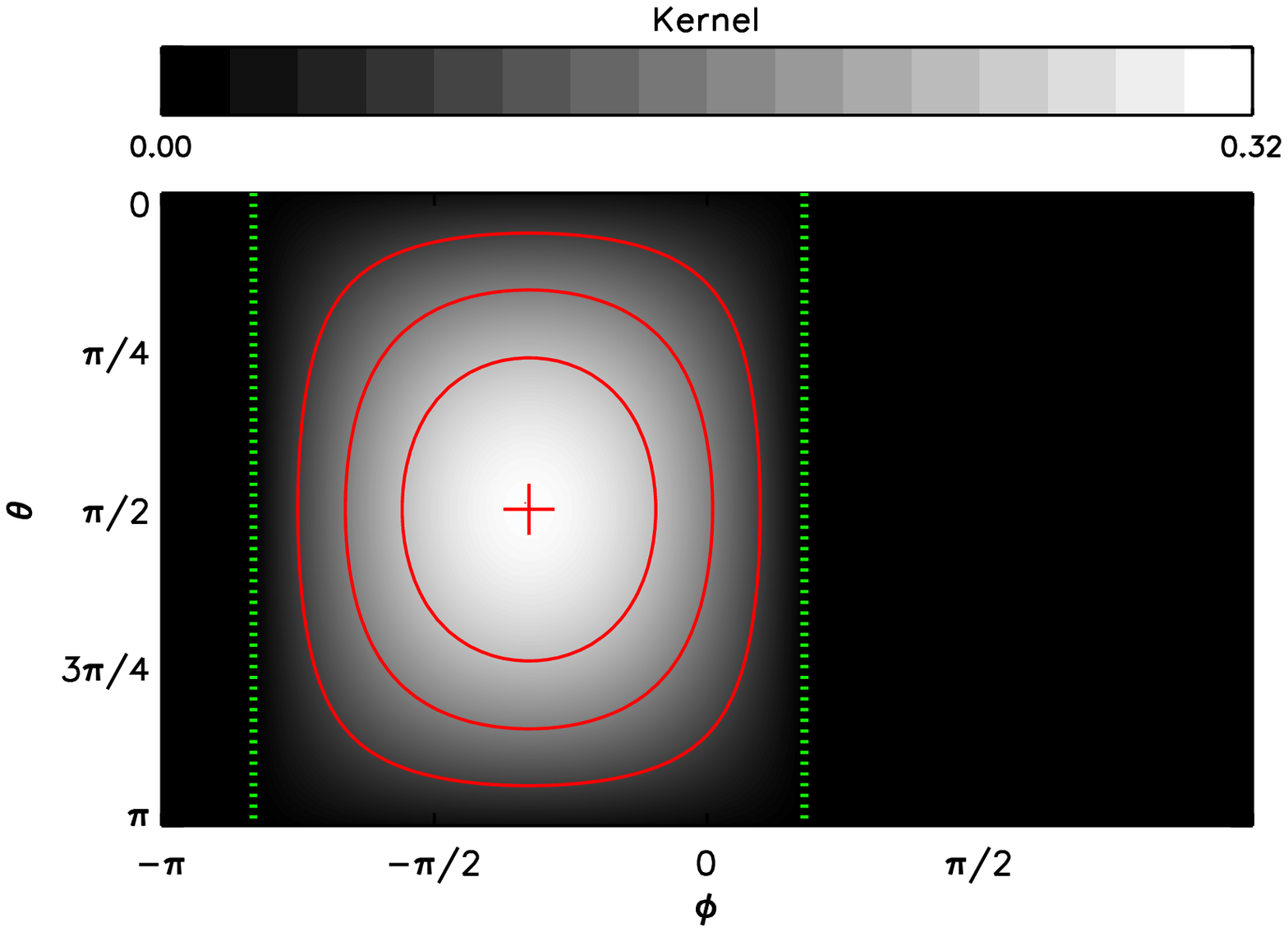}\includegraphics[width=85mm]{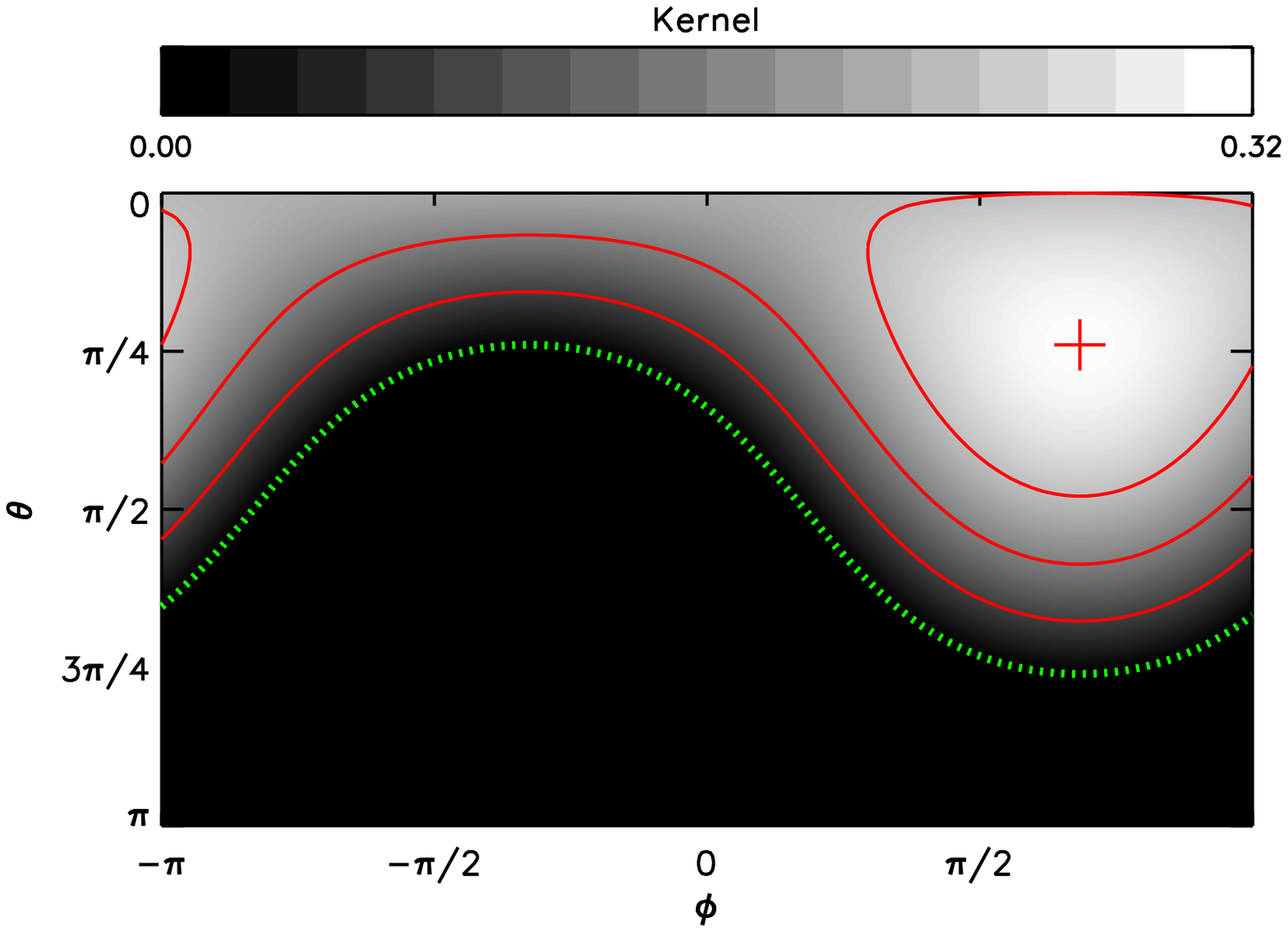}
\caption{The visibility (red contours) of a planet seen by an equatorial (left) or northern observer (right). The gray-scale shows the kernel of the convolution, which in this case is simply the rescaled visibility. The sub-observer location is denoted by a red cross.  The limb is shown in each case with green dotted lines. For either viewing geometry, the limb is a great circle and the non-zero portion of the kernel is a hemisphere.}
\label{vis_map}
\end{figure}

We first consider a planet viewed equator-on (left panel of Figure~\ref{vis_map}), which allows us to separate \eqref{thermal_harmonic_lightcurve} into two single integrals:
\begin{equation}\label{equatorial_thermal_equ}
F_{l}^{m}(t) =  \frac{N_l^m}{\pi} \int_{-1}^{1}\sqrt{1-x^2}P_{lm}(x) dx \int_{\phi_o-\frac{\pi}{2}}^{\phi_o+\frac{\pi}{2}} \cos(\phi-\phi_o) \cos(m\phi) d\phi,
 \end{equation}
where we have made the change of coordinates $x = \cos\theta$.  Note that we have given the example for a cosine $Y_l^m$ ($m\ge0$), but the sine instance ($m<0$) can be trivially obtained by replacing $m \to |m|$ and $\cos(m\phi) \to \sin(|m|\phi)$.  

The time-dependence can be factored out of the integral by making the change of variables $\Phi = \phi-\phi_o$, using a trigonometric identity, and noting that one of the two resulting integrals is zero (or simply using complex exponentials): 
\begin{equation}\label{equ_therm_sub}
F_{l}^{m}(t) =  \frac{N_l^m}{\pi} \cos(m\phi_o) \int_{-1}^{1}\sqrt{1-x^2}P_{lm}(x) dx \int_{-\frac{\pi}{2}}^{\frac{\pi}{2}} \cos(\Phi) \cos(m\Phi) d\Phi,
 \end{equation}
where the product of integrals is now solely a function of $l$ and $m$. This justifies the use of sinusoidal basis maps and lightcurves in the analysis of hot Jupiter thermal phase variations \citep{Cowan_2008}.
 
The integral of the associate Legendre polynomial is simplified by using a recurrence relation, then solved directly following \cite{Jepsen_1955}, as described in Appendix~A: 
 \begin{equation}\label{theta_integral}
\int_{-1}^{1}\sqrt{1-x^2}P_{lm}(x) dx =  \frac{1}{(2l+1)} \left[ R_{l+1}^{m+1} - R_{l-1}^{m+1} \right].
 \end{equation}
The integral of an associated Legendre polynomial over the interval $x \in [-1,1]$, represented here as $R_{l}^{m}$, depends on the parity of the spherical harmonic. If $l$ and $m$ are even, then $l-1$, $l+1$, and $m+1$ are odd; if $l$ and $m$ are odd, then $l-1$, $l+1$, and $m+1$ are even; if $l+m$ is odd, then so are $(l-1)+(m+1)$ and $(l+1)+(m+1)$.  Lastly, $R_{l-1}^{m+1} = R_{l+1}^{m+1}$ for odd $l>1$, producing a nullspace, $F_l^m=0$.

The $\Phi$-integral yields:
 \begin{equation}\label{phi_integral}
 \int_{-\frac{\pi}{2}}^{\frac{\pi}{2}} \cos\Phi \cos(m\Phi) d\Phi = \left\{ \begin{array}{ll} 
 \frac{\pi}{2}  & \textrm{if $|m|=1$}\\
 \frac{2}{1-m^2} \cos\left(\frac{m\pi}{2}\right) & \textrm{if $|m|\ne1$,} \end{array}\right.
 \end{equation}
which recovers the \cite{Cowan_2008} result of zero phase signature for odd $|m|>1$, since $\cos\left(\frac{m\pi}{2}\right)$ is zero in those cases. Physically, the brightness inhomogeneities cancel each other in the disk-integrated case. Mathematically, $Y_l^m$ with odd $|m|>1$ are in the nullspace of the convolution.

Combining \eqref{theta_integral} and \eqref{phi_integral} yields the full solution,
\begin{equation}\label{thermal_edge_on_solution}
F_{l}^{m}(t) = \left\{ \begin{array}{ll}
1 & \textrm{if $l=0$}\\
\frac{2}{\sqrt{3}} \cos\phi_o & \textrm{if $l=1$ and $m=1$}\\
\frac{2 (-1)^{m/2}}{\pi(1-m^2)}\sqrt{\frac{2(l-m)!}{(2l+1)(l+m)!}} \left[ R_{l+1}^{m+1}({\rm odd}) - R_{l-1}^{m+1}({\rm odd}) \right]  \cos(m\phi_o) & \textrm{if $l$ and $m$ are even}\\
0 & \textrm{otherwise,} \end{array}\right.
 \end{equation}
where $R_{l}^{m}({\rm odd})$ is given in Appendix~A. The nullspace is the union of odd $m>1$ ($\Phi$-integral goes to zero) and odd $l>1$ ($x$-integral goes to zero).

The first few non-zero harmonic lightcurves are:

\begin{minipage}[t]{0.45\linewidth}
\begin{align}
F_0^0(t) = &1\\
F_1^{1}(t)=&\frac{2}{\sqrt{3}}\cos\phi_o\\
F_2^{0}(t)=&-\frac{\sqrt{10}}{8}\\
F_2^{2}(t)=&\frac{\sqrt{15}}{8}\cos(2\phi_o)
\end{align}
\end{minipage}
\hspace{0.5cm}
\begin{minipage}[t]{0.45\linewidth}
\begin{align}
F_4^0(t)=&-\frac{3\sqrt{2}}{64}\\
F_4^2(t)=&\frac{\sqrt{5}}{32}\cos(2\phi_o)\\
F_4^4(t)=&-\frac{\sqrt{35}}{64}\cos(4\phi_o),
\end{align}
\end{minipage}
where again we stress that the sine harmonic lightcurves ($m<0$) can be trivially obtained by the substitution $\cos(m\phi_o) \to \sin(|m|\phi_o)$.

\subsection{Inclined Thermal Lightcurve}\label{inclined_thermal_section}
For a non-equatorial observer, one of the poles is visible and the other is invisible, rather than both being on the limb (right panel of Figure~\ref{vis_map}). 
The $\phi$-integral therefore runs from 0 to $2\pi$, while the $\theta$-integral runs from 0 to $\theta_{\rm limb}(\phi)$ if the north pole is visible, or from $\theta_{\rm limb}(\phi)$ to $\pi$ if the south pole is visible. With no loss of generality we consider a northern observer:
\begin{align}
F_{l}^m(t) =& \frac{N_l^m}{\pi} \left[ \sin\theta_{o} \int_{0}^{2\pi}\cos(\phi-\phi_o) \cos(m\phi) \int_{x_{\rm limb}}^{1}\sqrt{1-x^2} P_{lm}(x) dx d\phi \right.\nonumber \\
  &  + \left. \cos\theta_o \int_{0}^{2\pi} \cos(m\phi) \int_{x_{\rm limb}}^{1} x P_{lm}(x) dx d\phi \right],
 \end{align}
where $x_{\rm limb} = \cos\theta_{\rm limb}$.

As with the equatorial geometry, the time-dependence may be factored out of the integral using complex exponentials:
\begin{align}\label{inclined_thermal_integral}
F_{l}^m(t) =& \frac{N_l^m}{\pi} \cos(m\phi_o) \left[ \sin\theta_{o} \int_{0}^{2\pi}\cos\Phi \cos(m\Phi) \int_{x_{\rm limb}}^{1}\sqrt{1-x^2} P_{lm}(x) dx d\Phi \right.\nonumber \\
  &  + \left. \cos\theta_o \int_{0}^{2\pi} \cos(m\Phi) \int_{x_{\rm limb}}^{1} x P_{lm}(x) dx d\Phi \right],
 \end{align}
where \eqref{limb} dictates that 
\begin{equation}\label{x_limb}
x_{\rm limb} =   \frac{-\tan\theta_o \cos\Phi}{\sqrt{1+\tan^2\theta_o \cos^2\Phi}}.
\end{equation}

Although it is not immediately obvious, the integrals in \eqref{inclined_thermal_integral} are always zero for odd $l>1$. Following \S4 and \S5 of \cite{Russell_1906}, one can rotate to a coordinate system where the pole is at the sub-observer point without affecting $l$, which is analogous to total angular momentum. In this frame, the old $P_{lm}$ of (25) become sums over $m'$ of the new $P_{lm'}$. The  limits of integration simplify to $\int_0^1 dx$ and $\int_0^{2\pi}d\Phi$, allowing us to separate the $x$ and $\Phi$ integrations. The $\Phi$-integrals are of the form $\int_0^{2\pi} e^{i (m\pm 1) \Phi} d\Phi$ and $\int_0^{2\pi} e^{i m \Phi} d\Phi$, which vanish except when $m=\mp 1$ and $m=0$, respectively. In those cases where the $\Phi$-integrals don't kill (25), the $x$-integrals do, as  $\int_0^1 \sqrt{1-x^2} P_{l (\mp1)} dx$ and $\int_0^1 x P_{l0}(x) dx$ both vanish for odd $l>1$. Note that $m$ is analogous to the $z$-component of angular momentum and is not conserved in the rotation, so this coordinate system is not particularly helpful for computing general $F_{l}^{m}$. 

As with the equatorial case, one can use recurrence relations to express the $x$-integrals in \eqref{inclined_thermal_integral} as sums of simple integrals of $P_{lm}(x)$. But while there are recurrence relations for the indefinite integral $\int P_{lm}(x) dx$ \citep{DiDonato_1982}, we were unable to develop such a relation for the harmonic lightcurves, $F_l^m(t)$.  Instead, we use the brute force approach of substituting specific associated Legendre polynomials into \eqref{inclined_thermal_integral} and analytically solving the double integral. 

\begin{figure}
\begin{center}
\includegraphics[width=80mm]{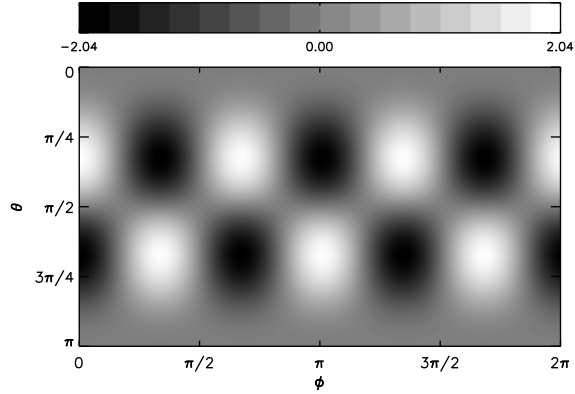}
\caption{The $Y_4^3$ brightness map is in the nullspace of the edge-on thermal lightcurve, but not of the inclined lightcurve. It is the lowest-order map that can produce power at odd $m>1$ in a thermal lightcurve.}
\label{ninja_map}
\end{center}
\end{figure}

We first solve the indefinite integral over $x$, which we evaluate at the limits of integration.  The resulting integrands for the $\Phi$-integrals include trigonometric functions with singularities. Since the sign of these functions can change on either side of the singularities, the integral over $\Phi$ exhibits unphysical jumps.\footnote{Although these jumps are undesirable, the derivative of the resulting curve is smooth, so \emph{Mathematica} is indeed returning a valid anti-derivative.}  Fortunately, the jumps occur at predictable fractions of $\pi$, and their amplitude is a tractable function of $\theta_o$, so the definite integral can be manually corrected.  The resulting low-order non-zero harmonic lightcurves are:  

\begin{minipage}[t]{0.35\linewidth}
\begin{align}
F_0^0 = & 1 \\
F_1^0 = & \frac{2\sqrt{6}}{3} \cos{\theta_o}\\
F_1^1 = & \frac{2}{\sqrt{3}} \sin{\theta_o}  \cos{\phi_o}\\
F_2^0 = & \frac{\sqrt{10}}{8} \left(3 \cos^2{\theta_o}-1 \right) \\
F_2^1 = & \frac{3}{4} \sqrt{\frac{5}{3}} \sin{\theta_o} \cos{\theta_o}  \cos{\phi_o}\\
F_2^2 = & \frac{\sqrt{15}}{8} \sin^2{\theta_o}  \cos(2 \phi_o)
\end{align}
\end{minipage}
\hspace{0.5cm}
\begin{minipage}[t]{0.55\linewidth}
\begin{align}
F_4^0= & \frac{-\sqrt{2}}{512}\left(9+20\cos(2\theta_o)+35\cos(4\theta_o) \right)\\
F_4^1= & \frac{5}{128\sqrt{10}} (24 \sin{ \theta_o}+2 \sin{ 2\theta_o } +7 \sin{ 4 \theta_o}) \cos{\phi_o}\\
F_4^2= & \frac{-\sqrt{5}}{64} (5+7 \cos{ 2 \theta_o} ) \sin^2{\theta_o} \cos(2 \phi_o)\\
F_4^3 = &\frac{35}{16\sqrt{70}}  \cos{  \theta_o}  \sin^3{\theta_o} \cos(3 \phi_o)\\
F_4^4 = & \frac{-\sqrt{35}}{64}    \sin^4{\theta_o} \cos(4 \phi_o).
\end{align}
\end{minipage}
The sine harmonic lightcurves ($m<0$) can be trivially obtained by the substitution $\cos(m\phi_o) \to \sin(|m|\phi_o)$.

The nullspace for an inclined geometry is more limited than the equatorial case: there are non-zero harmonic lightcurves with odd $m>1$ (provided that $l$ is even, e.g. $Y_4^3$ shown in Figure~\ref{ninja_map}).  

\section{Reflected Lightcurves}\label{reflected_lightcurves}
\subsection{Model Formalism}
We assume a spherical planet with a static albedo map on a circular orbit. For rotational lightcurves, the map only needs to be constant over a single rotation; or equivalently the recovered map is a diurnal average \citep[][]{Cowan_2009}. For rotational+orbital mapping \citep[``spin-orbit tomography,''][]{Fujii_2012}, the map is assumed to be static over an entire planetary orbit \citep[e.g., the Mars map of][]{Hasinoff_2011}. 

The reflection is treated as diffuse (Lambertian). Note, however, that real surfaces can exhibit specular reflection \citep[][]{Williams_2008} and atmospheres can exhibit Rayleigh or Mie scattering \cite[][]{Robinson_2010}.  The albedo map should be thought of as a top-of-atmosphere planetary albedo. 

For a uniform planet the resulting phase variations under the assumption of diffuse reflection is the well-known Lambert phase \citep{Russell_1916}.  In this section we derive higher moments of the lightcurve. This exercise is complementary to computing the phase variations of uniform but non-Lambertian planets \citep[e.g.,][]{Tousey_1957, Madhusudhan_2012}. Real planets are neither uniform nor Lambertian.
 
Albedo will in general be a function of wavelength, but since we consider only scattered light, there is no mixing of wavelengths. Our results can be generalized to any number of wavebands \citep[][]{Kawahara_2011, Fujii_2012} or arbitrary linear combinations of wavebands \citep[][]{Cowan_2009, Cowan_2011, Kawahara_2010, Cowan_2013}. 

The reflected-light kernel is $K(\theta, \phi, t) = \frac{1}{\pi}V(\theta, \phi, t) I(\theta, \phi, t)$, where the visibility, $V$ is defined as in \eqref{V}, while the illumination, $I$, is unity at the sub-stellar location, drops as the cosine of the angle from the sub-stellar location, $\gamma_s$, and is zero on the night-side of the planet:
\begin{equation}\label{I}
I(\theta, \phi, t) = \max[\cos\gamma_s, 0] = \max[\sin\theta\sin\theta_{s}\cos(\phi-\phi_{s})+\cos\theta\cos\theta_{s}, 0],
\end{equation}
and $\theta_s$ and $\phi_s$ are the sub-stellar co-latitude and longitude, respectively. Note that the kernel is proportional to the normalized weight, $W$, but is not divided by the Lambert phase function \citep[$W\equiv VI/\oint VId\Omega$,][]{Cowan_2011}. 

The reflected-light forward problem is therefore:
\begin{equation}
F_l^m(t) = \frac{1}{\pi}\oint V(\theta, \phi, t) I(\theta, \phi, t) Y_l^m(\theta, \phi) d\Omega,
\end{equation}
where $F$ in this case is the reflectance of the planet, i.e. the planet/star contrast ratio after accounting for the radius and semi-major axis of the planet, which are sometimes not of immediate interest. 

Noting that the orbital phase, $\alpha \in [0,\pi]$ ($\alpha=0$ at full phase; $\alpha=\pi$ at new phase), is simply the angular distance between the sub-observer and sub-stellar points, it may be expressed as
\begin{equation}\label{phase}
\cos\alpha = \sin\theta_{o}\sin\theta_{s}\cos(\phi_{o}-\phi_{s})+\cos\theta_{o}\cos\theta_{s} = \cos\xi \sin i,
\end{equation}
where the constant $i \in [0,\pi/2]$ is the orbital inclination with respect to the celestial plane ($i=0$ for a face-on orbit; $i=\pi/2$ for edge-on), and $\xi = \xi(0) + \omega_{\rm orb} t$ is the planet's true anomaly such that $\xi \in [0,2\pi]$, $\xi=0$ at superior conjunction, and $\xi=\pi$ at inferior conjunction.

As with the thermal emission problem, the crux stems from the piece-wise defined kernel, or equivalently, the limits of integration for the non-zero portion of the kernel.  The reflected lightcurve calculation is harder than the thermal lightcurve because the non-zero region of the integral is a lune rather than a hemisphere. 

The analytic expressions for the limb and terminator allow us to drop the piecewise-defined version of the visibility and illumination functions.  If one only considers those regions where both illumination and visibility are greater than zero, then the kernel is:
\begin{equation}
K_{nz}(\theta , \phi , t)  = \frac{1}{\pi}(\sin \theta \sin \theta_{o} \cos ( \phi - \phi_{o}) + \cos \theta \cos \theta_{o} ) (\sin \theta \sin \theta_{s} \cos ( \phi - \phi_{s}) + \cos \theta \cos \theta_{s} ).
\end{equation}

In the current study, we only consider the tidally locked configuration, which is likely to be relevant for the current crop of hot Jupiters as well as temperate planets orbiting low-mass stars, provided they are on circular orbits. The sub-stellar location on a tidally-locked planet is equatorial, $\theta_{s} = \pi/2$, $\theta_{o} = i$, and one can place the prime meridian at the sub-stellar meridian with no loss of generality, $\phi_{s} = 0$, so the kernel is simply $K_{nz}(\theta , \phi , t)  =  \frac{1}{\pi} (\sin i \sin^2\theta \cos\phi \cos(\phi-\phi_o) + \cos i \sin\theta\cos\theta\cos\phi).$ 

If the planet also orbits edge-on, $\theta_{o} = i = \frac{\pi}{2}$, then $K_{nz}(\theta , \phi , t)  =    \frac{1}{\pi}\sin^2 \theta \cos\phi \cos(\phi-\phi_o).$

\subsection{Tidally-Locked, Edge-On Reflected Lightcurve}
\begin{figure}
\includegraphics[width=85mm]{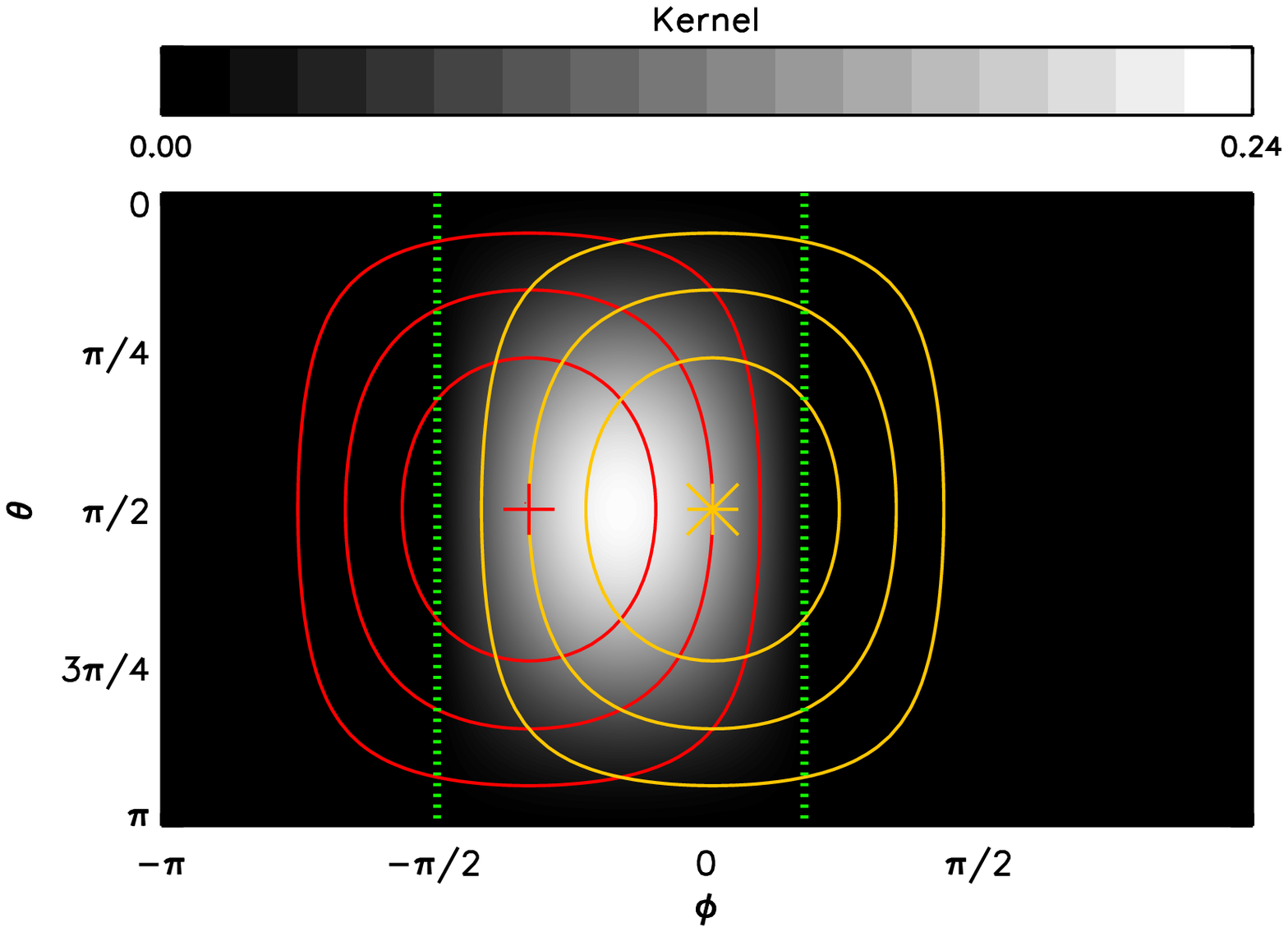}\includegraphics[width=85mm]{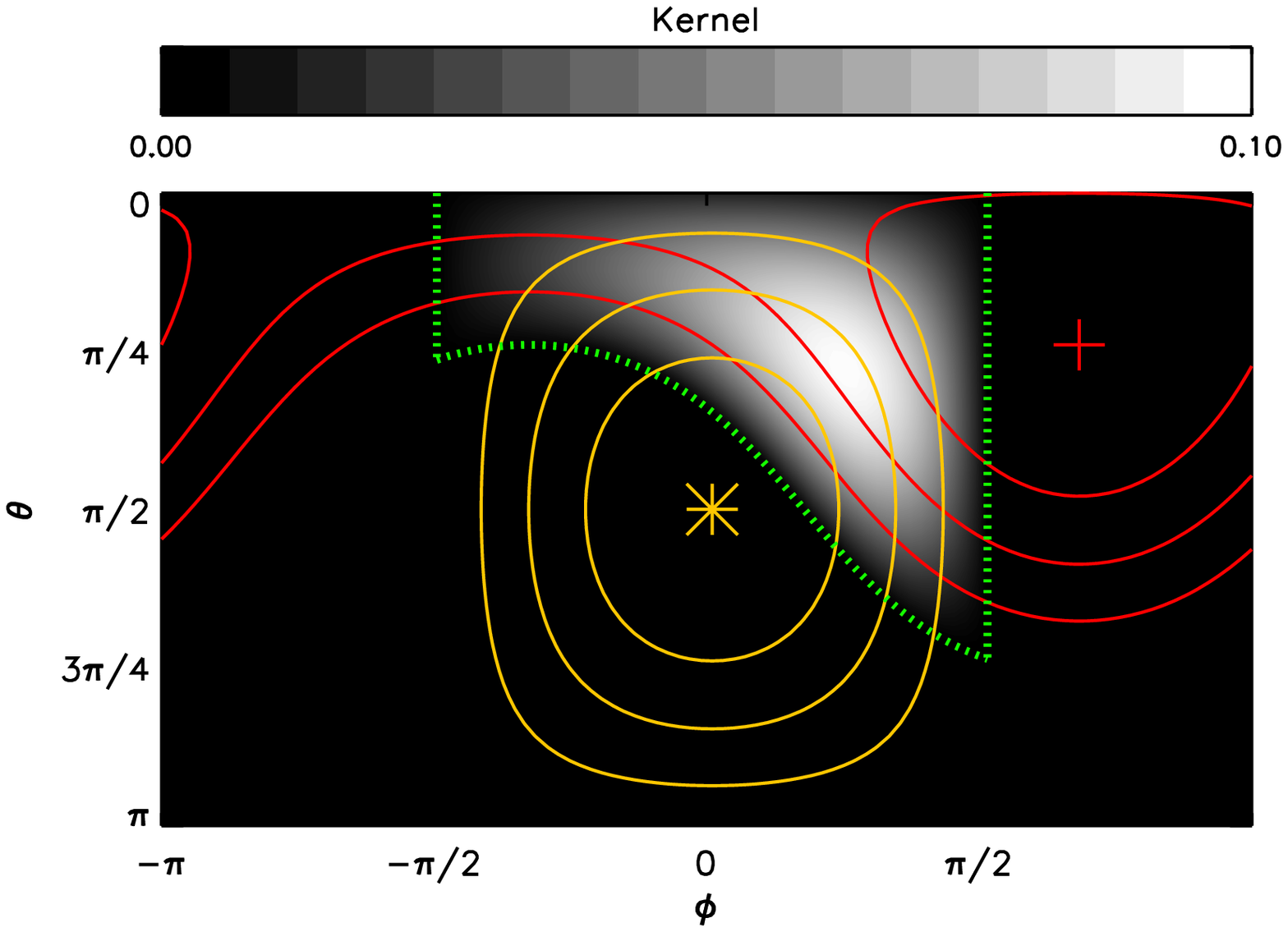}
\caption{The visibility (red contours) and illumination (yellow contours) of a tidally-locked planet seen by an equatorial (left) and northern (right) observer. The limb and terminator are shown in each case with green dotted lines. The sub-observer location is denoted by a red cross, the sub-stellar location by a yellow asterisk. The limb and terminator are both great circles, making the non-zero portion of the kernel a lune. The maximum value of the kernel, and its integral, are a function of the angular distance between the sub-observer and sub-stellar locations, i.e.: the planet's orbital phase.}
\label{ill_vis_map}
\end{figure}

The combination of tidally-locked rotation and edge-on orbit dictates that the $\theta$-integral runs from 0 to $\pi$, so the double integral can be split into two single integrals:
\begin{equation}
F_l^m(t)  =  \frac{N_l^m}{\pi} \int_{-1}^{1}(1-x^2)P_{lm}(x) dx  \int_{\phi_1}^{\phi_2} \cos\phi \cos(\phi-\phi_o) \cos(m\phi) d\phi,
\end{equation} 
where $\phi_1 = \max[-\pi/2, \phi_{o}-\pi/2]$, and $\phi_2 = \min[\pi/2, \phi_{o}+\pi/2]$ (left panel of Figure~\ref{ill_vis_map}), and we have again shown the cosine case. The $x$-integral is merely a scalar, so $F_l^m \propto F_\lambda^m$, as for the edge-on thermal case. The time-dependence cannot be factored out of the $\phi$-integral, however, so the negative and positive $m$ lightcurves are not trivially related by the substitution $\cos(m\phi) \to \sin(|m|\phi)$ (cf. the $m=-1$ and $m=1$ cases of Equation~\ref{edge_on_reflected_phi_integral}). 

As with the edge-on thermal lightcurve, the integral over $x$ may be solved for arbitrary $l$ and $m$ by using recurrence relations for associated Legendre polynomials and the \cite{Jepsen_1955} solution to their definite integral.  In the present case, this leads to terribly cumbersome expressions, so we instead adopt the brute force approach of analytically integrating the $x$-integral for specific $l$ and $m$.    

The $\phi$-integral must be solved separately in two cases: $\phi_o<0$ and $\phi_o \ge 0$ for the first and second halves of the planet's orbit, respectively. The two cases may be stitched together by noting that $\alpha = |\phi_o|$:
\begin{equation}\label{edge_on_reflected_phi_integral}
\int_{\phi_1}^{\phi_2} \cdots d\phi  = \left\{ \begin{array}{ll} \frac{-\sin(m\phi_o/2)}{m(m^2-4)}\Big( (m+2) \sin\left(\alpha - \frac{m\alpha}{2} +\frac{m\pi}{2}\right) + (m-2) \sin\left(\alpha + \frac{m\alpha}{2} -\frac{m\pi}{2}\right)\Big) & \textrm{if $m<-2$}\\[6pt]
\frac{1}{4}\sin\phi_o(\pi-\alpha+\sin\alpha\cos\alpha) & \textrm{if $m=-2$}\\[6pt]
\frac{1}{3}\sin\phi_o(1+\cos\alpha) & \textrm{if $m=-1$}\\[6pt]
\frac{1}{2}\Big(\sin\alpha+(\pi-\alpha)\cos\alpha\Big) & \textrm{if $m=0$}\\[6pt]
\frac{4}{3}\cos^4(\phi_o/2) & \textrm{if $m=1$}\\[6pt]
\frac{1}{4}\cos\phi_o(\pi-\alpha+\sin\alpha\cos\alpha) & \textrm{if $m=2$}\\[6pt]
\frac{\cos(m\phi_o/2)}{m(m^2-4)}\Big( (m+2) \sin\left(\alpha - \frac{m\alpha}{2} +\frac{m\pi}{2}\right) + (m-2) \sin\left(\alpha + \frac{m\alpha}{2} -\frac{m\pi}{2}\right)\Big) & \textrm{if $m>2$}. \end{array} \right.
\end{equation}

The first few non-zero harmonic lightcurves are:

\begin{minipage}[t]{0.40\linewidth}
\begin{align}
F_0^0(t) = & \frac{2}{3\pi}\Big( \sin\alpha + (\pi-\alpha)\cos\alpha\Big) \\
F_1^{-1}(t) = & \frac{\sqrt{3}}{8} \sin{\phi_o} (1+\cos{\phi_o})\\
F_1^1(t) = & \frac{\sqrt{3}}{2} \cos^4(\phi_o/2)\\
F_2^{-2}(t)=& \frac{\sin\phi_o}{\pi\sqrt{15}}  \Big( (\pi-\alpha)+\cos\alpha\sin\alpha \Big)\\
F_2^0(t) = & \frac{-2}{3\pi}\sqrt{\frac{2}{5}} \Big( \sin\alpha + (\pi-\alpha)\cos\alpha\Big)\\
F_2^2(t) = & \frac{2\cos\phi_o}{\pi\sqrt{15}}  \Big( (\pi-\alpha)+\cos\alpha\sin\alpha \Big)
\end{align}
\end{minipage}
\hspace{0.5cm}
\begin{minipage}[t]{0.50\linewidth}
\begin{align}
F_3^{-3}(t)=& \frac{1}{192}\sqrt{\frac{35}{2}} \Big(6\sin\phi_o+5\sin(2 \phi_o)-\sin(4 \phi_o)\Big)\\
F_3^{-1}(t)=&\frac{-1}{32} \sqrt{\frac{7}{6}} \sin\phi_o \Big(1+\cos\phi_o\Big)\\
F_3^1(t)=&\frac{-1}{8} \sqrt{\frac{7}{6}} \cos ^4\left(\phi_o/2\right)\\
F_3^3(t)=&\frac{1}{192} \sqrt{\frac{35}{2}} \Big(4\cos\phi_o + 5\cos(2\phi_o)-\cos(4\phi_o)\Big)\\
F_4^{-4}(t)=&\frac{4}{\pi \sqrt{35}} \frac{\sin^5\alpha \cos\alpha}{\sin\phi_o}\\
F_4^4(t)=&\frac{2}{\pi \sqrt{35}} \sin^3\alpha \cos(2\phi_o).
\end{align}
\end{minipage}

\begin{figure}
\begin{center}
\includegraphics[width=120mm]{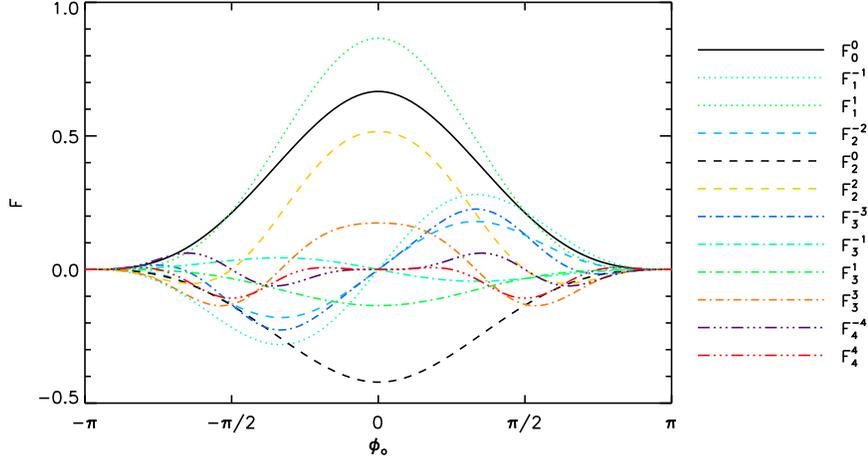}
\caption{Harmonic reflected lightcurves for a tidally-locked planet on an edge-on orbit. Line style denotes $l$, color denotes $m$.}
\label{edge_on_reflected}
\end{center}
\end{figure}

A uniform map, $M(\theta,\phi) \equiv 1$, produces the Lambert phase function, as expected.   The nullspace of the convolution includes odd $l-m$ (an odd number of nodes in the meridional direction), for which the contribution from the northern and southern hemispheres cancel perfectly. The harmonic lightcurves are shown in Figure~\ref{edge_on_reflected}. As expected, the negative and positive $m$ are not related by a simple phase shift.

Significantly, harmonic lightcurves with the same $m$ are proportional to each other, and $F_0^0$ is not flat.  As a result, and rather counterintuitively, the purely meridional map $Y_2^0(\theta,\phi)$ is not in the nullspace.

\subsection{Tidally-Locked, Inclined Reflected Lightcurve}
\begin{figure}
\begin{center}
\includegraphics[width=160mm]{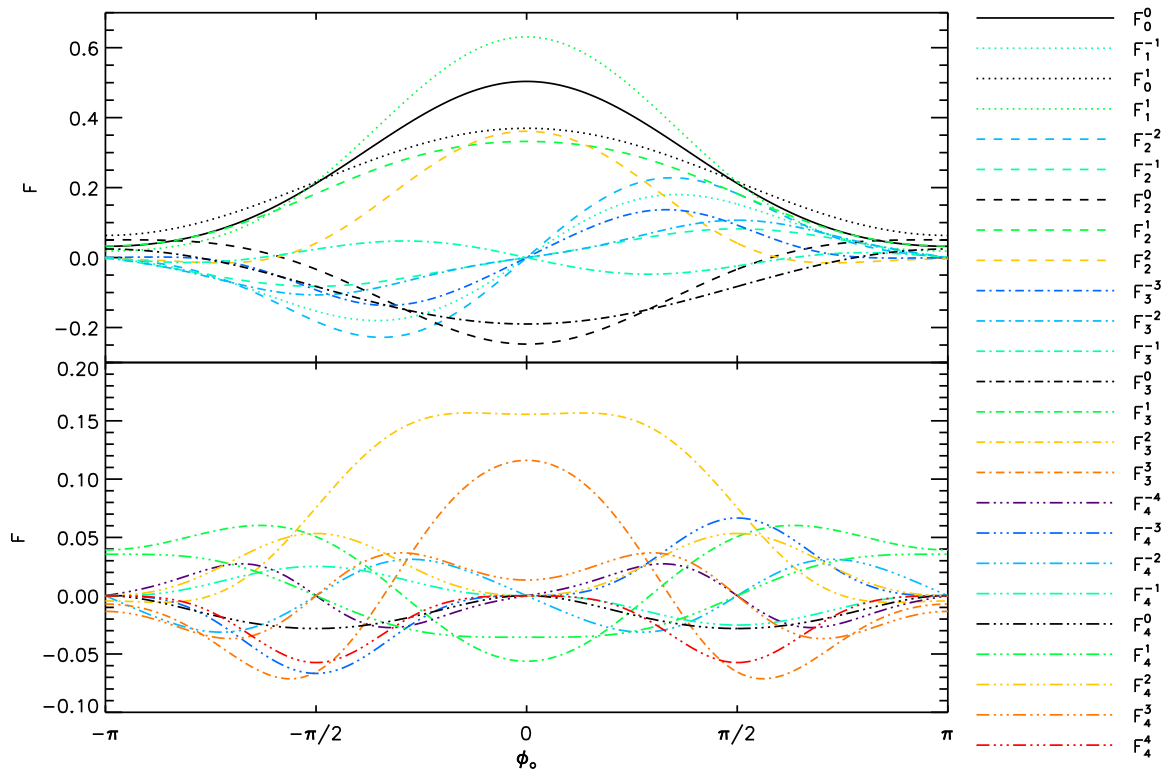}
\caption{Harmonic reflected lightcurves for a tidally locked planet on an inclined orbit ($i=\theta_o=\frac{\pi}{4}$). Line style denotes $l$, color denotes $m$.}
\label{general_reflected}
\end{center}
\end{figure}

In the more general case of a tidally-locked planet on an inclined orbit, the expression for harmonic lightcurves becomes:
\begin{align}
F_l^m(t)  =  \frac{N_l^m}{\pi} \Bigg\{ &\sin i \int_{-\pi/2}^{\pi/2} \cos\phi \cos(\phi-\phi_o) e^{im\phi} \left[\int_{x_{\rm limb}}^1(1-x^2)P_{lm}(x) dx \right] d\phi \nonumber\\
& + \cos i \int_{-\pi/2}^{\pi/2} \cos\phi e^{im\phi} \left[ \int_{x_{\rm limb}}^1x\sqrt{1-x^2}P_{lm}(x) dx \right] d\phi \Bigg\},
\end{align}
where $x_{\rm limb}$ is given by \eqref{x_limb} and the meridional limits of integration implicitly assume a northern observer (right panel of Figure~\ref{ill_vis_map}). For a southern observer the limits would be $[-1, x_{\rm limb}]$.  The limits of integration for the $\phi$-integral no longer span $2\pi$ radians and therefore do not lend themselves to the Fourier strategy used for the inclined thermal lightcurves.\footnote{Moreover, the locations of the jumps in the integrand of the $\phi$-integral are not as predictable as they are for inclined thermal lightcurves.}  Instead, we solve the integrals numerically and plot the solutions in Figure~\ref{general_reflected}.  

From numerical integration, we determine that there is no nullspace for the inclined reflected lightcurves up to $l=4$, nor do the lightcurves depend simply on $\theta_o$. In other words, one cannot simply scale the edge-on solutions of Figure~\ref{edge_on_reflected} by $\sin i$. For example, the shape of $F_2^0$ remains fixed but with the addition of a constant offset.  This means that the $Y_2^0$ component of the map has the effect of decreasing the reflectance of the planet at full phase and increasing it at crescent phases. This is essentially a restatement of the latitude-albedo effect noted by \cite{Cowan_2012b}: reflective poles make a low-obliquity planet appear abnormally bright at crescent phases.

\section{Discussion}\label{discussion}
\subsection{Degeneracies in Rotational Mapping}
The essential challenges of mapping distant bodies from time-resolved observations have been known for over a century \citep{Russell_1906}. Nullspaces are bigger in cases where the kernel has a fixed shape and latitude: the worst cases are thermal lightcurves and reflected rotational lightcurves of zero-obliquity objects at fixed phase. The nullspace is more limited for reflected phase variations (changing kernel shape) or reflected lightcurves of oblique rotators (changing kernel latitude).  Occultation mapping provides a much more varied kernel and is therefore nearly devoid of a nullspace. 

The problem of nullspaces for rotational lightcurves cannot be swept away by clever
parametrization.  For any planetary map that fits the data, one can add an
arbitrary linear combination of nullspace maps to obtain a very different map that
fits the data equally well.  This is true regardless of how the initial map
was parametrized. 

The only way to constrain the presence of the nullspace
maps is by adding \emph{a priori} constraints.  One universal constraint is that the map must be everywhere greater than zero, while albedo maps must additionally be less than unity everywhere on the planet \citep[these constraints are critical for rotational unmixing;][]{Cowan_2013}. The application of Tikhonov or maximum entropy regularization may help produce unique maps \citep[e.g.,][]{Donati_1997, Knutson_2007, Lanza_2009, Kawahara_2011}, but the validity of such additional constraints must be evaluated on a case-by-case basis. For example, the assumption of bimodal intensity may be reasonable for star spots, but is still being tested for the cloud-related markings on brown dwarfs.

Although these degeneracies make it difficult/impossible to obtain an accurate map of an unresolved body, it is still possible to precisely measure certain properties of a body based solely on rotational lightcurves.  

\subsection{Inclination-Dependent Nullspace}
\begin{figure}
\begin{center}
\includegraphics[width=120mm]{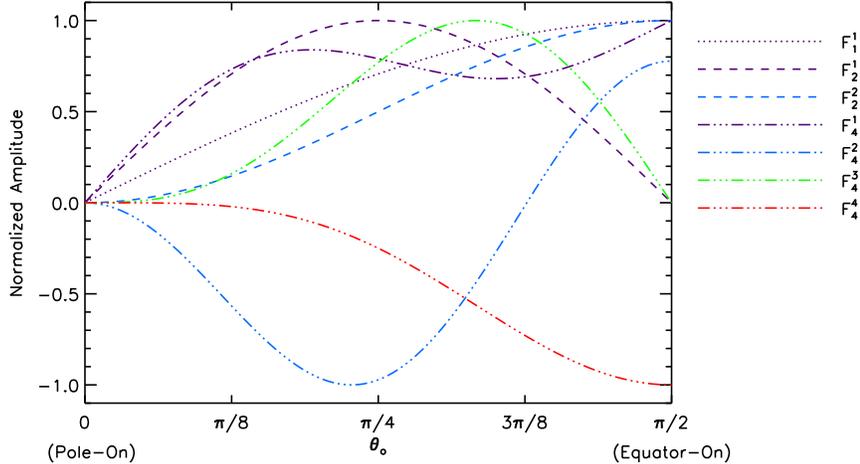}
\caption{The relative amplitude of harmonic thermal lightcurves as a function of sub-observer latitude (or, equivalently, rotational inclination).  Line style denotes $l$, color denotes $m$. The nullspace of the convolution is a function of $\theta_o$, allowing for qualitative estimates of rotational inclination based on the Fourier spectrum of observed lightcurves.}
\label{thermal_inclination}
\end{center}
\end{figure}

In Figure~\ref{thermal_inclination} we show the amplitude of low-order thermal harmonic lightcurves as a function of sub-observer latitude, $\theta_o$ ($\theta_o=0$ for pole-on or face-on rotation; $\theta_o=\pi/2$ for equator-on or edge-on rotation). The nullspace of the convolution and the amplitude of non-zero harmonic lightcurves are a function of $\theta_o$. A pole-on object obviously exhibits no rotational variability (left side of the plot). Moreover, an object exhibiting lightcurve power at $m=3$ is neither pole-on nor equator-on: the amplitude of $F_4^3$ exhibits a clear peak at $\theta_o=\pi/3$.   

Formally, there are an infinite number of harmonic maps that contribute to the lightcurve power at a given $m$:  
\begin{equation}
F^m(t) = \sum_{l=|m|}^{\infty} C_l^m F_l^m(t),
\end{equation}
where the coefficients $C_l^m$ are given by \eqref{coefficients} and the harmonic lightcurves are given by \eqref{harmonic_lightcurve}. Note that $F^m(t)$ is a simple sinusoid that can be extracted from an observed lightcurve via Fourier analysis. 

At first blush, the infinite series seems to scuttle attempts to attribute power in a particular Fourier mode to any one harmonic lightcurve. In practice, however, the low-pass nature of the convolution ensures that most of the power at a given $m$ comes from the lowest-order $Y_l^m$. Following the argument of Section~\ref{inclined_thermal_section}, the integral $\int_{0}^{1} x P_{l}(x) dx$ is $\{\frac{1}{2}, \frac{1}{3}, \frac{1}{8}, \frac{1}{48}, \frac{1}{128}\}$ for $l=\{0,1,2,4,6\}$, and in general $F_l^m \propto \Big(\Gamma \left(\frac{3}{2}-\frac{l}{2}\right) \Gamma \left(2+\frac{l}{2}\right)\Big)^{-1}$, where $\Gamma$ is the generalized factorial. Since high-$l$ harmonics are suppressed, it is reasonable to assume that the lightcurve power at some $m$ is primarily due to the $l=m$ component of the map if $m$ is one or even, or the $l=m+1$ component of the map if $m>1$ is odd.

If the intrinsic power in the $Y_4^3$ map was known \emph{a priori} and there was no other harmonic contributing $m=3$ power, then one might hope to estimate rotational inclination simply by measuring the $m=3$ power present in the lightcurve. But there would still be a two-way degeneracy because $\theta_o$ is double-valued for a given amplitude (green line in Figure~\ref{thermal_inclination}).

\subsection{Thermal Phase Variations of Short Period Planets}
It is significant that there are different nullspaces for the edge-on and inclined thermal lightcurves. One can't simply scale the edge-on solution of \cite{Cowan_2008} by $\sin i$. For transiting systems, the orbital inclination is nearly $90^\circ$ and is well measured by the transit morphology \citep[e.g.,][]{Charbonneau_2000}. This means that while the amplitude of odd $F^m(t)$ in an observed thermal lightcurve is likely an order of magnitude weaker than for adjacent even $F^m(t)$, a measurement of these modes could be converted into an estimate of the N-S asymmetry of the planet: lightcurve modes with odd $m>1$ must originate from N-S asymmetric spherical harmonics, because $l$ is even and $l-m$ is the number of meridional nodes. 

Moreover, the amplitude of odd modes in the observed lightcurve of a non-transiting planet provides a qualitative estimate of orbital inclination, since the planet may be assumed to have zero obliquity. The original motivation for measuring thermal phase variations of hot Jupiters was to estimate orbital inclination in order to break the $M\sin i$ degeneracy \citep{Agol_2005}.  That proposal was based on the presumption that all hot Jupiters would have the same day-night temperature contrast ($m=1$ amplitude).  While that assumption is demonstrably wrong \citep{Cowan_2007}, the inclination-dependance of thermal phase variation Fourier spectra offers an opportunity to constrain orbital inclination after all.

The best-characterized hot Jupiter, HD~189733b, only has empirical constraints up to $m=l=2$ at 3.6 and 4.5 micron \citep{Knutson_2012}. The amplitude of the $m=2$ component of the phase variations is 5\% of that at $m=1$ for both wavebands, indicating that the $m=2$ map has only 12\% the amplitude of the $m=1$ map \citep[following][]{Cowan_2008}.  This is not surprising since diurnal forcing primarily excites the $m=1$ mode.  The 8 micron 2D map of HD~189733b only constrains lightcurves to $l=1$ \citep{Majeau_2012}, so the strength of odd $m>1$ modes is currently unknown empirically. The N--S asymmetry that could result in odd lightcurve modes is debated theoretically \citep[][]{Cho_2003,Cooper_2005}. Note that this asymmetry is in principle accessible to eclipse mapping \citep[][]{Majeau_2012, deWit_2012}. Aside from the intrinsically weak signal, possible complications include the presence of eccentricity seasons \citep{Lewis_2013}, the contamination of $m=2$ modes by ellipsoidal variations \citep{Cowan_2012}, and limb-darkening, which slightly modifies the convolution kernel and may admit non-zero harmonic lightcurves for odd $l>1$ \citep{Russell_1906, Cowan_2008}. 

\subsection{Rotational Inclination of Stars and Brown Dwarfs}
Likewise, the Fourier spectra of spotty stars \citep{Lanza_2009} and brown dwarfs \citep{Artigau_2009, Radigan_2012} might hint at their rotational inclination.  This is unsurprising given that single-band star spot modeling can yield inclination estimates good to tens of degrees \citep[][]{Walker_2007}, despite myriad degeneracies \citep[][]{Dorren_1987, Kipping_2012}.\footnote{Note that multi-band photometry can partially break these degeneracies because of the wavelength-dependence of spot/photosphere contrast and limb-darkening \citep{Harmon_2000}.} It should be feasible, by the same token, to use measurements of odd modes in Spitzer Space Telescope lightcurves of spotty brown dwarfs to obtain a qualitative estimate of their inclination.

\begin{figure}
\begin{center}
\includegraphics[width=120mm]{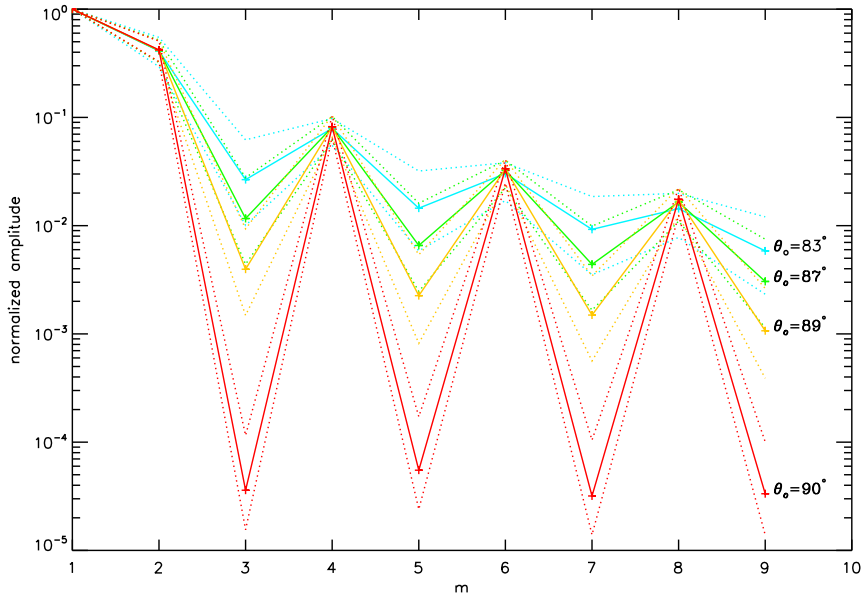}
\caption{Fourier spectra for 1000 stars with randomly generated spot maps, viewed at 4 different inclinations. The spectra have been normalized to their $m=1$ amplitude. The solid lines show the median Fourier spectrum for each inclination, while the dotted lines denote the $1\sigma$ confidence intervals.}
\label{spot_test}
\end{center}
\end{figure}

The stellar/brown dwarf inclination inverse problem is more favorable than extracting the orbital inclination of hot Jupiters because the signal-to-noise ratio for the variations are much greater (typically $\sim 1$\% for an active star, rather than $\sim0.1$\% for a hot Jupiter), and the intrinsic power spectrum of the thermal map does not drop precipitously with $m$.  The limiting case of a single $\delta$-function spot has a flat Fourier spectrum. This makes Fourier analysis an inefficient means of mapping star spots, but potentially useful for constraining rotational inclination.   

In Figure~\ref{spot_test} we show a Monte Carlo simulation of power spectra for 1000 rotating stars at 4 different inclinations.  The maps have a background photospheric intensity of unity and each map has between 1 and 3 randomly located square spots of zero intensity.  The sizes of the spots are normally distributed about 0.2~radians, with a standard deviation of 0.1~radians (we take the absolute value of negative values). For each spot map, we generate simulated lightcurves for observers at a variety of inclinations, neglecting limb darkening, using a $400\times200$ spatial grid, and adopting a temporal resolution of 200. We then compute the Fourier spectrum of each lightcurve and normalize it to the $m=1$ amplitude.  The solid lines show the median Fourier spectra for each inclination, while the dotted lines denote the $1\sigma$ confidence intervals.  

Edge-on rotators, shown by red lines in Figure~\ref{spot_test},  produce no power in the odd modes; the non-zero values on the plot are due to numerical errors \citep[see also][]{Aigrain_2012}. For stars that are rotating nearly edge-on, the power at odd $m$ is a very strong function of orientation.  A factor-of-two measurement of odd $F^m(t)$ can constrain rotational inclination to $\pm 6^\circ$ (cf. the lines for $83^\circ$ and $89^\circ$ in Figure~\ref{spot_test}). Note that this test is conservative since we are considering \emph{only} the $m=3$ amplitude.  The inclination estimates could be improved by measuring more odd modes or by additional constraints such as map positivity.  

While the \emph{a priori} odds of a star being within $10^{\circ}$ of edge-on are not great, they increase substantially for stars known to host transiting planets \citep{Sanchis_2012}.  At smaller inclinations (more face-on), the orientation of the star still leaves an imprint on its Fourier spectrum, but it is not clear whether a Fourier approach would offer any advantage over the usual star spot modeling in such a case.   

The intrinsic power in the $Y_4^3$ map will depend on the latitude of star-spots, with equatorial spots providing no power at this harmonic.  But star spots are, if anything, less likely to be at the equator \citep[][]{Baumann_2004}. In any case, the Monte Carlo varies the spot map, so our accuracy estimate accounts for the ability of a particular map to mimic a different inclination. If the actual distribution of spots is not random, then the accuracy of such estimates may improve or deteriorate.  For example, if all stars have perfectly N-S symmetric intensity maps, then they will never exhibit odd $F^m(t)$, and odd modes will not be useful indicators of inclination \citep[this seems unlikely given the complexity of spot maps:][]{Donati_1997, Silva_2010}. It is not yet known whether the clear/cloudy regions on L/T transition dwarfs have preferred latitudes, let alone whether they are N--S symmetric.

Fourier analysis of rotating inhomogeneous bodies is similar, conceptually and mathematically, to estimating stellar inclinations with asteroseismology  \citep{Gizon_2003}. Although our proposed method requires high-precision photometry, the observational cadence could be considerably lower than for asteroseismology, so many \emph{Kepler} target stars may be amenable to such an analysis, without the need for $v\sin i$ measurements from high-resolution stellar spectra \citep[e.g.,][]{Hirano_2012}. As with any star-spot based inclination estimate, problems include differential rotation of star spots, or their formation/dissipation on rotational timescales \citep{Silva_2011}. If the brightness markings evolve on timescales longer than the rotational period, it should still be possible to Fourier decompose each rotation separately to perform the spectral power analysis.   Limb-darkening is again a source of systematic uncertainty/error.  

\subsection{Obliquity of Directly-Imaged Planets}
Directly imaged jovian planets emit thermal radiation because they are still young.  Although no rotational variability has yet been reported for these objects, they are probably cloudy \citep[e.g.,][]{Madhusudhan_2011} and may exhibit the same sort of variability as brown dwarfs.  Merely detecting this rotational modulation would be a technical feat with important implications for giant planet formation.  Further down the road, however, we might hope to measure their rotational lightcurves with sufficient precision to constrain their rotational inclination with respect to our line of sight. Unlike hot Jupiters, directly imaged jovian planets are not tidally locked, but their orbital inclination may be estimated observationally.  The Fourier spectrum of such a rotational lightcurve would therefore put a joint constraint on the planet's obliquity and equinox via:
\begin{equation}\label{obliquity_equation}
\cos\theta_{o} = \sin\Theta\cos\xi_{\Theta}\sin i + \cos\Theta\cos i,
\end{equation}
where $\Theta$ is the planetary obliquity and $\xi_{\Theta}$ is the angular location of northern summer solstice with respect to superior conjunction. Fortuitously, polarimetry of planetary thermal emission can constrain the sky-plane components of a planet's rotation axis \citep[][]{deKok_2011}, enabling full determination of a planet's spin axis.

Other techniques have been proposed to measure the obliquity of directly-imaged planets.  Full-orbit thermal phase curves of mature planets, for which insolation rather than internal heat dominates the power budget, might betray the obliquity of a planet, but the inverse problem is complicated by orbital eccentricity, diurnal heating, and the details of heat storage and transport \citep{Gaidos_2004, Cowan_2012c}. Full-orbit reflected lightcurves of directly-imaged planets have also been demonstrated to convey information about planetary obliquity \citep{Fujii_2012}.  These methods have the advantage that they can break the degeneracy between obliquity and its orientation, but they require observations spanning the planetary orbit rather than its rotation, and therefore may not be practical for long-period planets.  

\subsection{Reflected Phase Variations of Planets}
The fact that $F_l^0 \propto F_0^0$ for edge-on reflected phase curves has important implications for the retrieval of albedo from planetary phase variations. The $Y_2^0$ map corresponds to a zonally uniform planet with bright poles and has the effect of reducing the amplitude of zeroth-order, Lambertian, phase variations (c.f.\ Equations~44 and 48). Consider the worst-case of a diffusely-reflecting planet with albedo map $M = \frac{1}{3}Y_0^0 + \frac{2}{3\sqrt{10}}Y_2^0$, which has an albedo of unity at both poles, zero at the equator, and a mean albedo of $\langle M\rangle = \frac{1}{3}$ (qualitatively similar to a planet with polar snow/ice).  The Bond albedo of the planet is $A = \frac{1}{3}A_0^0 + \frac{2}{3\sqrt{10}}A_2^0 = \frac{1}{4}$ (Appendix~\ref{bond_albedo}).  The reflected lightcurve for the planet, however, is $F(t) = \frac{1}{3}F_0^0(t) + \frac{2}{3\sqrt{10}}F_2^0(t) = \frac{1}{5}F_0^0(t)$.  In other words, an observer would see a planet exhibiting perfectly Lambertian phase variations with an apparent albedo of $A^*(t) \equiv F(t)/F_0^0(t)=0.2$. The sub-observer and sub-stellar latitudes are both equatorial, but the albedo estimate differs from the actual Bond albedo by $0.05/0.2=25$\% (and differs from the mean albedo by $67\%$).  One should therefore be wary of estimating an exoplanet's Bond albedo, even if the planet orbits edge-on, is tidally locked, and exhibits Lambertian phase variations.  Precision is no guarantee of accuracy.

The inclined tidally-locked reflected lightcurves exhibit no nullspace up to $l=4$, but the harmonic lightcurves are not orthogonal, leading to formal degeneracies for mapping planets, even in the limit of noiseless data.  It remains to be seen to what extent such degeneracies affect the spin-orbit exo-cartography of \cite{Fujii_2012}: in that more general case the changing shape and latitude of the kernel should produce a much more limited nullspace.  Although we have only derived reflected lightcurves for tidally-locked planets, it should still be the case that meridional albedo markings affect reflected phase variations, provided one averages over the rotational variation.  Moreover, the numerical simulations presented in \cite{Cowan_2012b} suggest that many of our results will carry over to small but non-zero obliquity.   
  
\section*{Acknowledgments}
We thank the anonymous referee for substantive comments that improved the manuscript. NBC thanks J.H.~Steffen and W.M.~Farr for useful Mathematica tips, and J.~de~Wit for thoughtful commentary. HMH gratefully acknowledges support from the National Science Foundation (NSF) International Research Fellowship Program (IRFP) under Grant No. OISE-1159218.

\appendix
 
\section[]{Integrals of Associated Legendre Polynomials}
A technical crux of the analytic forward problem is solving definite integrals of associated Legendre polynomials, $P_{lm}$.

\subsection{Recurrence Relation}
Recurrence relations allow us to relate functions of $x$ including associated Legendre polynomials to combinations of simple $P_{lm}$. For example, we used
\begin{equation}
\sqrt{1-x^2} P_{lm} =  \frac{-1}{2l+1} \left[ P_{l-1,m+1} - P_{l+1,m+1} \right].
\end{equation}

\subsection{Definite Integrals on $x \in [-1,1]$}
For the special cases where the limits of integration are $x \in [-1,1]$, compact solutions (i.e., not involving sums) have been worked out by \cite{Jepsen_1955}.  Those authors solved the definite integral of $P_{lm}(x)$ without the Condon-Shortley phase, precisely what we need in the current paper: 
\begin{equation}\label{Jepsen_Eqn}
R_l^m \equiv \int_{-1}^{1} P_{lm}(x) dx  =  \left\{ \begin{array}{cr} R_l^m(\textrm{even}) \equiv \frac{2m[(l/2)!]^2(l+m)!}{l[(l-m)/2]! [(l+m)/2]! (l+1)!} & \textrm{if $l$ and $m$ are even}\\[6pt]
	 R_l^m(\textrm{odd}) \equiv-\frac{\pi m(l+m)! (l+1)!}{l 2^{2l+1} \{[(l+1)/2]!\}^2 [(l-m)/2]! [(l+m)/2]!} & \textrm{if $l$ and $m$ are odd}\\[6pt]
	 0 & \textrm{if $l+m$ is odd,} \end{array} \right.
\end{equation}
Physically, the northern and southern hemispheres have perfectly canceling lightcurves in the third case. 

\section[]{Bond Albedo of a Tidally-Locked Planet}\label{bond_albedo}
The Bond albedo for a tidally-locked planet is time-invariable:
\begin{equation}
A = \frac{1}{\pi}\oint A(\theta, \phi) I(\theta, \phi, t) d\Omega.
\end{equation}

The contribution to the Bond albedo from a cosine harmonic map is
\begin{equation}
A_l^m = \frac{N_l^m}{\pi} \int_{-1}^{1}\sqrt{1-x^2}P_{lm}(x) dx \int_{-\frac{\pi}{2}}^{\frac{\pi}{2}} \cos\phi \cos(m\phi) d\phi,
\end{equation}
which is identical to \eqref{equ_therm_sub} but with $\phi_o \equiv 0$. By analogy, the solution is simply
\begin{equation}\label{edge_on_Bond_albedo}
A_{l}^{m} = \left\{ \begin{array}{ll}
1 & \textrm{if $l=0$}\\[6pt]
\frac{2}{\sqrt{3}} & \textrm{if $l=1$ and $|m|=1$}\\[6pt]
\frac{2 (-1)^{m/2}}{\pi(1-m^2)}\sqrt{\frac{2(l-m)!}{(2l+1)(l+m)!}} \left[ R_{l+1}^{m+1}({\rm odd}) - R_{l-1}^{m+1}({\rm odd}) \right]  & \textrm{if $l$ and $m$ are even}\\[6pt]
0 & \textrm{otherwise,} \end{array}\right.
 \end{equation}
where $R_{l}^{m}({\rm odd})$ is given by \eqref{Jepsen_Eqn}. The nullspace is the union of odd $|m|>1$ ($\phi$-integral goes to zero) and odd $l>1$ ($x$-integral goes to zero).

\end{onecolumn}

\begin{thebibliography}{99}
\bibitem[Agol 
\& Charbonneau(2005)]{Agol_2005} Agol, E., \& Charbonneau, D.\ 2005, Spitzer Proposal, 20482

\bibitem[Aigrain et al.(2012)]{Aigrain_2012} Aigrain, S., Pont, F., 
\& Zucker, S.\ 2012, MNRAS, 419, 3147

\bibitem[Artigau et al.(2009)]{Artigau_2009} Artigau, {\'E}., 
Bouchard, S., Doyon, R., \& Lafreni{\`e}re, D.\ 2009, ApJ, 701, 1534

\bibitem[Aster, Borchers \& Thurber (2013)]{Aster_2013} Aster, R.C., Borchers, B.\ and Thurber, C.H., 2013, Parameter Estimation And Inverse Problems (Elsevier Academic Press) 

\bibitem[Baumann et 
al.(2004)]{Baumann_2004} Baumann, I., Schmitt, D., Sch{\"u}ssler, M., \& Solanki, S.~K.\ 2004, A\&A, 426, 1075

\bibitem[Charbonneau et al.(2000)]{Charbonneau_2000} Charbonneau, D., 
Brown, T.~M., Latham, D.~W., \& Mayor, M.\ 2000, ApJ, 529, L45

\bibitem[Cho et al.(2003)]{Cho_2003} Cho, J.~Y.-K., Menou, K., 
Hansen, B.~M.~S., \& Seager, S.\ 2003, ApJ, 587, L117

\bibitem[Cooper 
\& Showman(2005)]{Cooper_2005} Cooper, C.~S., \& Showman, A.~P.\ 2005, ApJ, 629, L45

\bibitem[Cowan et al.(2007)]{Cowan_2007} Cowan, N.~B., Agol, E., 
\& Charbonneau, D.\ 2007, MNRAS, 379, 641

\bibitem[Cowan \& Agol(2008)]{Cowan_2008} Cowan, N.~B., \& Agol, E.\ 2008, ApJ, 678, L129 

\bibitem[Cowan et al.(2009)]{Cowan_2009} Cowan, N.~B., Agol, E., 
Meadows, V.~S., et al.\ 2009, ApJ, 700, 915 

\bibitem[Cowan et al.(2011)]{Cowan_2011} Cowan, N.~B., Robinson, 
T., Livengood, T.~A., et al.\ 2011, ApJ, 731, 76 

\bibitem[Cowan et al.(2012a)]{Cowan_2012} Cowan, N.~B., Machalek, 
P., Croll, B., et al.\ 2012, ApJ, 747, 82

\bibitem[Cowan et al.(2012b)]{Cowan_2012b} Cowan, N.~B., Abbot, 
D.~S., \& Voigt, A.\ 2012, ApJ, 752, L3

\bibitem[Cowan et al.(2012c)]{Cowan_2012c} Cowan, N.~B., Voigt, A., 
\& Abbot, D.~S.\ 2012, ApJ, 757, 80

\bibitem[Cowan 
\& Strait(2013)]{Cowan_2013} Cowan, N.~B., \& Strait, T.~E.\ 2013, ApJ, 765, L17

\bibitem[de Kok et al.(2011)]{deKok_2011} de Kok, R.~J., Stam, 
D.~M., \& Karalidi, T.\ 2011, ApJ, 741, 59

\bibitem[Deutsch(1958)]{Deutsch_1958} Deutsch, A.~J.\ 1958, 
Electromagnetic Phenomena in Cosmical Physics, 6, 209

\bibitem[Deutsch(1970)]{Deutsch_1970} Deutsch, A.~J.\ 1970, ApJ, 
159, 985 

\bibitem[DiDonato(1982)]{DiDonato_1982} DiDonato, A.R.\ 1982, Mathematics of Computation, 38, 547

\bibitem[Donati 
\& Collier Cameron(1997)]{Donati_1997} Donati, J.-F., \& Collier Cameron, A.\ 1997, MNRAS, 291, 1 

\bibitem[Dorren(1987)]{Dorren_1987} Dorren, J.~D.\ 1987, ApJ, 320, 
756

\bibitem[Fujii et al.(2011)]{Fujii_2011} Fujii, Y., Kawahara, H., 
Suto, Y., et al.\ 2011, ApJ, 738, 184 

\bibitem[Fujii 
\& Kawahara(2012)]{Fujii_2012} Fujii, Y., \& Kawahara, H.\ 2012, ApJ, 755, 101

\bibitem[Gaidos 
\& Williams(2004)]{Gaidos_2004} Gaidos, E., \& Williams, D.~M.\ 2004, New Astronomy, 10, 67 

\bibitem[Gizon \& Solanki(2003)]{Gizon_2003} Gizon, L., \& Solanki, S.~K.\ 2003, ApJ, 589, 1009 

\bibitem[Harmon 
\& Crews(2000)]{Harmon_2000} Harmon, R.~O., \& Crews, L.~J.\ 2000, AJ, 120, 3274 

\bibitem[Hasinoff et al.(2011)]{Hasinoff_2011} Hasinoff, S.W., Levin, A., Goode, P.R., Freeman, W.T., 2011 IEEE Internat.\ Conf.\ on Computer Vision (ICCV), 185

\bibitem[Hirano et al.(2012)]{Hirano_2012} Hirano, T., 
Sanchis-Ojeda, R., Takeda, Y., et al.\ 2012, ApJ, 756, 66

\bibitem[Jepsen et al.(1955)]{Jepsen_1955} Jepsen, D.W., Haugh, E.F. \& Hirschfelder, J.O.\ 1955, Proc Natl Acad Sci,  41(9), 645-7

\bibitem[Kawahara 
\& Fujii(2010)]{Kawahara_2010} Kawahara, H., \& Fujii, Y.\ 2010, ApJ, 720, 1333

\bibitem[Kawahara 
\& Fujii(2011)]{Kawahara_2011} Kawahara, H., \& Fujii, Y.\ 2011, ApJ, 739, L62

\bibitem[Kipping(2012)]{Kipping_2012} Kipping, D.~M.\ 2012, MNRAS, 
427, 2487

\bibitem[Knutson et al.(2007)]{Knutson_2007} Knutson, H.~A., 
Charbonneau, D., Allen, L.~E., et al.\ 2007, Nature, 447, 183

\bibitem[Knutson et al.(2012)]{Knutson_2012} Knutson, H.~A., Lewis, 
N., Fortney, J.~J., et al.\ 2012, ApJ, 754, 22

\bibitem[Lacis 
\& Fix(1972)]{Lacis_1972} Lacis, A.~A., \& Fix, J.~D.\ 1972, ApJ, 174, 449

\bibitem[Lanza et 
al.(2009)]{Lanza_2009} Lanza, A.~F., Pagano, I., Leto, G., et al.\ 2009, A\&A, 493, 193

\bibitem[Lewis et al.(2013)]{Lewis_2013} Lewis, N.~K., Knutson, 
H.~A., Showman, A.~P., et al.\ 2013, ApJ, 766, 95

\bibitem[Madhusudhan et al.(2011)]{Madhusudhan_2011} Madhusudhan, N., 
Burrows, A., \& Currie, T.\ 2011, ApJ, 737, 34

\bibitem[Madhusudhan 
\& Burrows(2012)]{Madhusudhan_2012} Madhusudhan, N., \& Burrows, A.\ 2012, ApJ, 747, 25

\bibitem[Majeau et al.(2012)]{Majeau_2012} Majeau, C., Agol, E., 
\& Cowan, N.~B.\ 2012, ApJ, 747, L20

\bibitem[Marcialis(1988)]{Marcialis_1988} Marcialis, R.~L.\ 1988, AJ, 
95, 941

\bibitem[Oakley \& Cash(2009)]{Oakley_2009} Oakley, P.~H.~H., \& Cash, W.\ 2009, ApJ, 700, 1428

\bibitem[Radigan et al.(2012)]{Radigan_2012} Radigan, J., 
Jayawardhana, R., Lafreni{\`e}re, D., et al.\ 2012, ApJ, 750, 105

\bibitem[Robinson et al.(2010)]{Robinson_2010} Robinson, T.~D., 
Meadows, V.~S., \& Crisp, D.\ 2010, ApJ, 721, L67 

\bibitem[Russell(1906)]{Russell_1906} Russell, H.~N.\ 1906, ApJ, 
24, 1

\bibitem[Russell(1916)]{Russell_1916} Russell, H.~N.\ 1916, ApJ, 
43, 173

\bibitem[Sanchis-Ojeda et al.(2012)]{Sanchis_2012} Sanchis-Ojeda, 
R., Fabrycky, D.~C., Winn, J.~N., et al.\ 2012, Nature, 487, 449

\bibitem[Silva-Valio et 
al.(2010)]{Silva_2010} Silva-Valio, A., Lanza, A.~F., Alonso, R., \& Barge, P.\ 2010, A\&A, 510, A25

\bibitem[Silva-Valio 
\& Lanza(2011)]{Silva_2011} Silva-Valio, A., \& Lanza, A.~F.\ 2011, A\&A, 529, A36

\bibitem[Tremaine(1991)]{Tremaine_1991} Tremaine, S.\ 1991, Icarus, 
89, 85

\bibitem[Tousey(1957)]{Tousey_1957} Tousey, R.\ 1957, Journal of 
the Optical Society of America (1917--1983), 47, 261

\bibitem[Walker et al.(2007)]{Walker_2007} Walker, G.~A.~H., Croll, 
B., Kuschnig, R., et al.\ 2007, ApJ, 659, 1611 

\bibitem[Wild(1991)]{Wild_1991} Wild, W.~J.\ 1991, ApJ, 368, 622

\bibitem[Williams 
\& Gaidos(2008)]{Williams_2008} Williams, D.~M., \& Gaidos, E.\ 2008, Icarus, 195, 927 

\bibitem[Winn et al.(2005)]{Winn_2005} Winn, J.~N., Noyes, R.~W., 
Holman, M.~J., et al.\ 2005, ApJ, 631, 1215

\bibitem[de Wit et al.(2012)]{deWit_2012} de Wit, J., Gillon, M., Demory, B.-O., \& Seager, S.\ 2012, A\&A, 548, A128

\end{thebibliography}
\end{document}